\documentclass[a4paper,12pt]{article}
\usepackage[pctex32]{graphics}

\begin{document}
\topmargin 0pt
\oddsidemargin 0mm

\newcommand{\alp}{\alpha}
\newcommand{\bta}{\beta}
\newcommand{\gmm}{\gamma}
\newcommand{\del}{\delta}
\newcommand{\omg}{\omega}
\newcommand{\sgm}{\sigma}
\newcommand{\lmd}{\lambda}
\newcommand{\tha}{\theta}
\newcommand{\vph}{\varphi}
\newcommand{\Alp}{\Alpha}
\newcommand{\Bta}{\Beta}
\newcommand{\Gmm}{\Gamma}
\newcommand{\Del}{\Delta}
\newcommand{\Omg}{\Omega}
\newcommand{\Sgm}{\Sigma}
\newcommand{\Lmd}{\Lambda}
\newcommand{\Tha}{\Theta}
\newcommand{\half}{\frac{1}{2}}
\newcommand{\rnd}{\partial}
\newcommand{\nab}{\nabla}

\newcommand{\beqn}{\begin{eqnarray}}
\newcommand{\eeqn}{\end{eqnarray}}
\newcommand{\be}{\begin{equation}}
\newcommand{\ee}{\end{equation}}

\begin{titlepage}

\vspace{5mm}
\begin{center}
{\Large \bf Does entropic force always imply the Newtonian force
law? } \vspace{12mm}

{\large   Yun Soo Myung \footnote{e-mail
 address: ysmyung@inje.ac.kr}}
 \\
\vspace{10mm} {\em  Institute of Basic Science and School of
Computer Aided Science, Inje University, Gimhae 621-749, Republic of
Korea}

\end{center}

\vspace{5mm} \centerline{{\bf{Abstract}}}
 \vspace{5mm}
We study the entropic force  by introducing a bound $S \le
A^{3/4}$ between entropy and area  which was derived by imposing
the non-gravitational collapse condition.  In this case, applying
a modified entropic force  to this system does not lead to the
Newtonian force law.

\end{titlepage}
\newpage
\renewcommand{\thefootnote}{\arabic{footnote}}
\setcounter{footnote}{0} \setcounter{page}{2}

\section{Introduction}
Recently, Verlinde has proposed the Newtonian force law as an
entropic force by using  the equipartition rule and the holographic
principle~\cite{Ver}. After  his work,  the dynamics of apparent
horizon in the Friedmann-Robertson-Walker universe~\cite{SG}, the
Friedmann equations~\cite{Pad1,CCO}, the connection in the loop
quantum gravity~\cite{Smo}, the accelerating
surfaces~\cite{makela1}, holographic actions for black hole
entropy~\cite{CM}, and application to holographic dark
energy~\cite{LWh} were considered from the entropic force.
Furthermore,  cosmological implications were reported in \cite{cos},
the extension to Coulomb force~\cite{Wang}, and the symmetry aspect
of entropy force~\cite{Zhao} were investigated. The entropic force
was also discussed in the black hole spacetimes~\cite{bhole,bh-MK}.
Its connection to the uncertainty principle was considered
in~\cite{UP}.

We briefly review what was going on the entropic force.
Explicitly, when a test particle with mass $m$ is close to a
surface (holographic screen) with distance $\Delta x$ (compared to
the Compton wave length $\lambda_m=\frac{\hbar}{mc}$), the change
of entropy on the holographic screen takes the form \be
\label{eq1} \Delta S=2\pi k_B\frac{\Delta x}{\lambda_m} \to  2\pi
m \Delta x \ee in the natural units of $\hbar=c=k_B=1$ and
$G=l^2_{pl}$. Considering that the entropy of a system depends on
the distance $\Delta x$, an entropic force $F$ could be arisen
from the thermodynamical conjugate of the distance \be \label{eq2}
F \Delta x=T \Delta S \ee which is considered as an indication
that the first law of thermodynamics may be  realized  on the
holographic screen. Plugging (\ref{eq1}) into (\ref{eq2}) leads to
a connection between entropic force and temperature \be
\label{eq3} F=2\pi m T \ee which implies that if one knows the
temperature $T$, the entropic force is determined by (\ref{eq3}).
In order to define the temperature $T$ on the screen, we  assume
that the energy $E$ is distributed on a spherical shape of
holographic screen with radius $R$ and  the mass $M$ is located at
the origin of coordinate as the source. Then, we introduce  the
holographic principle, the equality of energy and mass, and the
equipartition rule~\cite{Pad2,Pad3}, respectively, as
\beqn \label{eq4n}N&=&\frac{A}{G} \\
\label{eq4m}E&=&M,\\
 \label{eq4e} E&=&\frac{1}{2 }N T \eeqn
with the area of the holographic screen $A=4\pi R^2$. These are
combined to determine the temperature on the screen \be
\label{eq5} T=\frac{GM}{2\pi R^2}. \ee Substituting (\ref{eq5})
into (\ref{eq3}),  the entropic force is realized as the Newtonian
force law \be \label{eq6} F=\frac{G m M}{R^2}. \ee

In this work, we use a ``modified entropic force" to derive the
Newtonian force law  by considering two  entropy bounds of $S\le
A$ and $S \le A^{3/4}$ instead of the  ``entropic force". The
former bound leads to the Newtonian force law, while the latter
does not provide the Newtonian force law.
 The latter
accounts for the ordinary matter which  is determined  by the
non-gravitational collapse condition.

\section{Two  issues on defining the temperature}

It is well known  that two unusual  assumptions to derive the
temperature were the holographic principle (\ref{eq4n}) and the
equipartition rule (\ref{eq4e}). Concerning the former, an urgent
issue is how one can construct a spherically holographic screen of
radius $R$ which encloses a source mass $M$ located at the origin
using the holographic principle.  This is an important
issue~\cite{bh-MK} because the holographic screen (an exotic
description of spacetime) originates from relativistic approaches
to black hole ~\cite{hoo,Sus} and cosmology~\cite{Bou,SL}.
Verlinde has introduced this screen  by analogy with an absorbing
process of a particle around the event horizon of black hole.
Considering a smaller test mass $m$ located at $\Delta x$ away
from the screen and getting the change of entropy on the  screen,
its behavior should resemble that of a particle approaching a
stretched horizon of a black hole, as was described by
Bekenstein~\cite{Bek}. It is clear  that Verlinde has introduced
the holographic screen as a basic input to derive the entropic
force.

The other issue is on the latter: why the equipartition rule could
be applied to this non-relativistic  surface to define the
temperature without any justifications.  For black holes, the
equipartition rule becomes the Smarr formula of $E=NT/2=2ST$ when
using the relation of $N=\frac{A}{G}=4S$. Also, it can be derived
from the first law of thermodynamics $dE=TdS$ for the
Schwarzschild black hole where the Komar charge is just the ADM
mass $M$.   Even though the equipartition rule may be  available
for the classical (thermodynamic) system, the holographic
principle of $N=A/G$ is not guaranteed to apply to any
non-relativistic situations. In this sense, this issue is closely
related to the first issue.

If the above two questions are answered properly, one would  make a
further step to understand the origin of Newtonian  force via the
entropic force.  However, there remains  a gap between
non-relativistic approach (absence of horizons) and relativistic
approach (presence of horizons).

\section{Modified entropic force}
In this section, we wish to develop  another issue of modified
entropic force.
 In deriving the
non-relativistic Newtonian force law (\ref{eq6}), we assume that the
surface is between the test mass $m$ and the source mass $M$, but
the test mass is very close to the surface as compared to its
Compton wavelength $\lambda_m$. According to Bekenstein's argument
in deriving the area quantum  of the Schwarzschild black
hole~\cite{Bek}, the test particle is indistinguishable from horizon
itself if  the test particle is on the order of Compton wavelength
away from the event horizon. That is, a relativistic quantum
particle cannot be localized to better than its Compton wavelength,
yielding a lower bound on the increment of the black hole horizon
area \be \label{area-q}\Delta A \ge (\Delta A)_{\rm min}=8 \pi l^2_p
\ee due to the assimilation of a neutral test particle. This
implies, in turn,
 the increment of black hole entropy \be \label{entinc}  \Delta S
\approx \Delta A. \ee In other words, considering the event horizon
as the holographic screen where relativistic effects dominate and
the Newtonian approach   is no longer a good scheme, the distance
$\Delta x$ of Compton wavelength is not a relevant requirement on
increasing the area of event horizon~\cite{Hod}.

 At this stage, we
remind the reader that the entropy increase (\ref{eq1}) was derived
from a simple analogy with entropic explanation of thermodynamically
emergent forces on polymers immersed in a heat bath~\cite{Ver}.
Therefore, it is not easy to see how the relation (\ref{eq1}) works
on the gravity side.

 Hence, we have to develop   another logic to  explain a
difference  between (\ref{eq1}) and (\ref{entinc}). To this end,
Smolin~\cite{Smo} has proposed that the information change in
entropy carries  the flux of a bit or byte across the surface
which is necessarily discretized. He has proposed such a process
that a small excitation initially in the interior region moves out
to the exterior, where it may be  interpreted as a massive
particle. Jacobson's idea~\cite{Jac} has shown  that any
translation of an excitation across the boundary surface involves
a change both of energy $U$ and entropy $S$, where the latter
implies a change of the area of the boundary. Further, Verlinde's
idea ~\cite{Ver} has implied that there must be a temperature $T$
associated with this process since any change $\Delta U$ in energy
is accompanied by $\Delta S$.  Smolin has proposed that the change
$\Delta U$ in energy corresponds to this translational motion over
a distance $\Delta x$. This  means that there exists a force
$F=\frac{\Delta U}{\Delta x}$ acting on the excitation.  According
to the first law of thermodynamics, this force takes the form \be
\label{f-eq} F \Delta x=\Delta U=T\Delta S\ee which implies {\it a
modified entropic force} \be \label{ff-eq}F=T\frac{\Delta
S}{\Delta x}. \ee Here, we mention that this modified equation
differs   from (\ref{eq2}), even though their forms are the
same~\cite{Smo,MR}.  Importantly, if  one gives up the linear
relation (\ref{eq1}) between $\Delta S$ and $\Delta x$, and then,
it may be  replaced by a  relationship between entropy $S$ and
area $A$ as
 \be S=S(A). \ee

\section{ $S \sim A$ versus $S \sim A^{3/4}$ }

From now on, we use the modified relation (\ref{ff-eq})  to study
the entropic force.   It is well known that the entropy of
Schwarzschild black hole is given by the Bekenstein-Hawking
entropy~\cite{Haw,Bek} \be \label{bh-ent}S_{BH}=\frac{A}{4l^2_p}.
\ee However, the nature of this entropy is one of the greatest
mysteries of modern physics because it scales as the area of black
hole rather than its volume. This peculiar property has led to the
holographic principle~\cite{hoo,Sus,Bou}, stating  that the number
of degrees of freedom in any system including gravity effects
grows only as the area of its boundary. If  the entropy of surface
is
 taken to be (\ref{bh-ent}), one finds
 \be \label{m-entr}
 \Delta S= \frac{\Delta A}{4 l^2_p}. \ee
It is noted  that in (\ref{ff-eq}), $\Delta S$ is one fundamental
unit of entropy when $\Delta x \simeq \lambda_m=1/m$ (if one
considers (\ref{area-q}) further, $\Delta S= 2\pi$) and the
entropy gradient points radially from outside of the boundary
surface to inside. We can easily check  that Eq.(\ref{ff-eq})
together with (\ref{m-entr}) leads to the Newtonian force law as
in (\ref{eq6}) when using the temperature (\ref{eq5}).

On the other hand, it is known that  gravitational collapse limits
the entropy of a physical system.  Information (entropy) requires
the energy, while formation of a horizon by gravitational collapse
restricts the amount of energy allowed in a finite region.
Explicitly, 't Hooft has shown that if one excludes configuration
whose energies are so large that they inevitably undergo
gravitational collapse, one finds the {\it non-covariant entropy
bound}~\cite{hoo} \be \label{non-cov}S \le
\Big(\frac{A}{l^2_p}\Big)^{3/4} \ee which is clearly different from
the {\it covariant entropy bound} \be \label{co-ent} S \le
S_{BH}\sim \frac{A}{l^2_p}. \ee As a concrete example, we consider a
thermal system of radius $R$ and temperature $T$ which implies that
its entropy (energy) are given by $S\sim T^3R^3(E\sim T^4R^3$).
Requiring the non-gravitational collapse condition \be
\label{grav-c} E < \frac{R}{l^2_p}, \ee one obtains the temperature
bound \be T<\frac{1}{\sqrt{l_pR}}. \ee Then, the entropy bound
appears as in (\ref{non-cov}) \be \label{ent-b}S
<\Big(\frac{R}{l_p}\Big)^{3/2} = \Big(\frac{A}{l^2_p}\Big)^{3/4}.
\ee

There are several arguments which support that the non-covariant
bound has  more application   than  the covariant entropy bound to a
system of  the ordinary matter.   The authors~\cite{CKN} have
proposed that the entropy bound (\ref{ent-b}) could be derived from
the energy bound of $E \le E_{BH}$, when getting rid of many states
where the Schwarzschild radius is much larger than the system size.
Hence the non-covariant entropy bound (\ref{ent-b}) is more
restrictive than the covariant entropy bound
(\ref{co-ent})~\cite{Myungent}.   Buniy and Hsu~\cite{BH} have shown
that the entanglement entropy has the same bound as in (\ref{ent-b})
when imposing the non-gravitational collapse condition
(\ref{grav-c}). Also, Chen and Xiao~\cite{CX} have proved  that
under the condition of (\ref{grav-c}), the entropy bound for the
local quantum field theory is $A^{3/4}$ but not $A$ for considering
either  bosonic fields or fermionic fields in the system of size
$l$. The authors~\cite{KLM} have confirmed  that under the condition
of (\ref{grav-c}), the entropy bound of the local quantum field
theory has taken to be (\ref{ent-b})  when using the generalized
uncertainty principle.   Additionally, considering the spacetime
foam uncertainty of $\delta l \ge l_{\rm
p}^{\alpha}l^{\alpha-1}$~\cite{Ng,MSeo}, it was shown that the case
of $\alpha=2/3$ could explain the holographic model with infinite
statistics whose entropy scales as $S\sim A$, while the case of
$\alpha=1/2$ could  describe the ordinary matter with Bose-Fermi
statistics whose entropy scales as  $S\sim A^{3/4}$. Recently,  the
reasonable arguments has been given that the correct bound should be
$S<A/4$, and not $S<A^{3/4}$ for the cosmological matter
distributions~\cite{BFL}. However, that work conjectured that the
stronger bound of $S <A^{3/4}$ does hold for static, weakly
gravitating systems. Here, we  consider the static, weakly
gravitating systems because we are working with the
non-gravitational collapse condition (\ref{grav-c}).

Finally, we would like to  mention the possibility that the
covariant entropy bound (\ref{co-ent}) works when it applies to
general relativity. In a modified gravity of $f(R)$
theory~\cite{NOr}, instead,  the entropy takes the
form~\cite{NOe,FT}
 \be
S_{f}=\frac{f'(R)A}{4l^2_p}. \ee Hence, the non-covariant entropy
bound of (\ref{non-cov}) may be constructed from $f(R)$ gravities.

 \section{Entropic force on the ordinary matter}

At this stage, we could not confirm that the non-covariant bound
(\ref{non-cov}) takes into account the  entropy of an ordinary
matter including weakly gravitating effects  exactly, instead of the
covariant entropy bound (\ref{co-ent}) for the strongly gravitating
systems of  black hole, de Sitter cosmological horizon and apparent
horizon in cosmology. However, there is no reason to prefer  the
maximum entropy of $S_{\rm max}\sim A$ to find the entropic force
between ordinary matters (source particle and test particle).  As
was previously mentioned, there are several theoretical evidences
which support that the bound (\ref{non-cov}) is better appropriate
for describing the ordinary matter than the bound (\ref{co-ent}).

Hence,  in this work,  we use the maximum entropy of $S_{\rm max}
\sim A^{3/4}$  in order to derive the entropy force for the system
because the source mass $M$ is not a black hole.  Considering
$S=\alpha (A/l^2_p)^{3/4}$, its variation is given by \be \Delta
S=\frac{\partial S}{\partial A}\Delta
A=\frac{3\alpha}{4l^{3/2}_p}\frac{\Delta A}{A^{1/4}}. \ee In this
case, the entropic force takes the form \be \label{oent-f} F=
T\frac{\Delta S}{\Delta x}=\frac{3\alpha
T}{4l^{3/2}_pA^{1/4}}\frac{\Delta A}{\Delta x} \ee which leads to
\be \label{non-N} F=G^{1/4}\frac{G mM}{R^{5/2}} \ee when choosing
the temperature $T$ in (\ref{eq6}) and a coefficient $\alpha$ as
\be \alpha=\frac{(4\pi)^{1/4}}{3}. \ee It is evident that
Eq.(\ref{non-N}) does not represent the Newtonian force law. In
order to derive the Newtonian force law, one has to explain why
the entropy of the surface (holographic screen) should be  given
by the area-law. We note that (\ref{oent-f}) was used to make a
correction to the Newtonian force law by considering the entropy
corrections~\cite{MR}.

 In addition to two issues of holographic
principle and equipartition theorem on determining the temperature
on the screen, there exists the third issue on a form of entropy
of the screen when using the modified entropic force. It is
clarified that if $S$ does not satisfy an area-law (for example,
$S \sim A^{3/4}$), one could not obtain the Newtonian force law
from the modified entropic force (\ref{ff-eq}).

\section{Discussions}

It is fare to say that the origin of the gravity is  not yet fully
understood. If the gravity is not a fundamental force, it may be
emergent from the other approach to gravity.  Newtonian force law
may be emergent from the equipartition rule and the holographic
principle~\cite{Ver}.

As was mentioned previously, it may  be not  proper to use the
linear relation (\ref{eq1}) between $\Delta S$ and $\Delta x$ on
the gravity side. According to Bekenstein~\cite{Bek},  a classical
point particle could not increase the area of black hole horizon.
On the other hand,  a relativistic quantum particle cannot be
localized to better than its Compton wavelength
$\lambda_m$~\cite{Hod}. This yields a lower bound on the increment
of the black hole area, due to the assimilation of a test
particle. This is regarded as the origin of the entropy increase
in the black hole.   In order to explain this gab, Smolin has
modified the linear relation (\ref{eq1}) into a relation
(\ref{m-entr}) between $\Delta S$ and $\Delta A$~\cite{Smo}.

Furthermore, we wish to point out  that the source mass $M$ behind
the screen is not a black hole. Also the surface is not the event
horizon of a black hole.  Hence, the area-law entropy of $S \sim
A$ is not justified to represent the entropy of ordinary matter.
For this purpose, we introduced   the non-covariant entropy of $S
\sim A^{3/4}$ obtained when imposing the non-gravitational
collapse condition (\ref{grav-c}) on the ordinary matter. Here, we
found the non-appearance of Newtonian force law.  In this sense,
the non-appearance of Newtonian force may be related to the fact
that entropy bound requests further modification.

Consequently, the appearance of Newtonian force law from the
entropic force (\ref{eq2}) seems not to be robust. The Newtonian
force law was obtained  when employing holographic principle and
equipartition theorem to derive the temperature  and using the
modified entropic force (\ref{ff-eq}) together with the area-law
entropy.  If one uses another entropy bound of $S\sim A^{3/4}$
together with (\ref{ff-eq}), however,  one fails to derive the
Newtonian force law.

\section*{Acknowledgment}
This work was   supported  by Basic Science Research Program through
the National Research  Foundation (NRF) of Korea funded by the
Ministry of Education, Science and Technology (2010-0028080).

\end{document}